\documentclass[twocolumn,showpacs]{revtex4}
\usepackage{times,xspace}
\usepackage{amsbsy,amssymb,amsmath,bm}
\usepackage{graphicx,color,epsfig,rotate}
\usepackage{fancyheadings}

\def\bbbc{{\mathchoice {\setbox0=\hbox{$\displaystyle\rm C$}\hbox{\hbox
to0pt{\kern0.4\wd0\vrule height0.9\ht0\hss}\box0}}
{\setbox0=\hbox{$\textstyle\rm C$}\hbox{\hbox
to0pt{\kern0.4\wd0\vrule height0.9\ht0\hss}\box0}}
{\setbox0=\hbox{$\scriptstyle\rm C$}\hbox{\hbox
to0pt{\kern0.4\wd0\vrule height0.9\ht0\hss}\box0}}
{\setbox0=\hbox{$\scriptscriptstyle\rm C$}\hbox{\hbox
to0pt{\kern0.4\wd0\vrule height0.9\ht0\hss}\box0}}}}

\newcommand{\ignore}[1]{}
\newcommand{\mComment}[1]{}
\newcommand{\gComment}[1]{}
\newcommand{\jComment}[1]{}
\newcommand{\rComment}[1]{}
\newcommand{\lComment}[1]{}

\renewcommand{\mComment}[1]{\textcolor{blue}{Manny: #1}}
\renewcommand{\gComment}[1]{\textcolor{red}{Gerardo: #1}}
\renewcommand{\jComment}[1]{\textcolor{green}{Jim: #1}}
\renewcommand{\rComment}[1]{\textcolor{magenta}{Ray: #1}}
\renewcommand{\lComment}[1]{\textcolor{purple}{Rolando: #1}}

\begin{document}

\title{Thermally Increasing Correlation/Modulation Lengths
and Other Selection Rules in Systems with Long Range
Interactions}
\author{Zohar Nussinov}
\address{The Department of Physics and Astronomy, The University of 
California at Los Angeles, CA 90095*}

\date{Received \today }

\begin{abstract}
In this article, addressing large $n$ systems, 
we report that in numerous systems
hosting long and short range interactions, 
multiple correlation lengths may appear. 
The largest correlation
lengths often monotonically increase with
temperature and diverge in the 
high temperature limit. 
Notwithstanding, the magnitude of the 
correlations themselves decreases 
with increasing temperature.
We examine correlation function 
in the presence of competing interactions
of long and short ranges. The behavior of the correlation
and modulation lengths as a function of temperature 
provides us with selection rules on the possible underlying
microscopic interactions. As a concrete 
example of these notions, we consider
the correlations in a system of screened
Coulomb interactions coexisting with 
attractive short range interactions.
 
\end{abstract}


\maketitle

\section{Introduction} 
Systems with long range interactions abound
in nature \cite{long}. Several examples in the current condensed matter
arena include Coulomb gases
(plasmas) which encompass- amongst
many others, the omnipresent free electron gas, quantum Hall
systems \cite{lilly}, \cite{QHE}, \cite{Fogler}, 
adatoms on metallic surfaces, amphiphilic systems \cite{amp}, 
interacting elastic defects
(dislocations and disclinations)
in solids \cite{vortex3}, interactions amongst vortices in 
fluid mechanics \cite{vortex1} and superconductors \cite{vortex2},
crumpled membrane systems \cite{Seul}, wave-particle interactions 
\cite{wavpart}, interactions amongst holes 
in cuprate superconductors \cite{steve}, \cite{us}, \cite{zohar},
\cite{Low}, \cite{carlson}, manganates and nickelates \cite{cheong}, 
\cite{golosov}, some theories of structural
glasses \cite{dk}, \cite{peter}, \cite{Gilles}, \cite{new},
and colloidal systems \cite{der}, \cite{reich}.
Needless to say, the list of systems 
(and works) goes far beyond
the little outlined here.
Much of the work to date focused on 
the character of the transitions 
in these systems and the subtle 
thermodynamics that is often 
observed (e.g., the equivalence
between different ensembles in many such
systems is no longer
as obvious, nor always correct,
as it is in the ``canonical'' 
short range case \cite{barre}). Other 
very interesting aspects of
different systems have been addressed
in \cite{azbel}. 

\begin{figure}[htb]
\vspace*{-0.5cm}
\includegraphics[angle=-90,width=7.6cm]{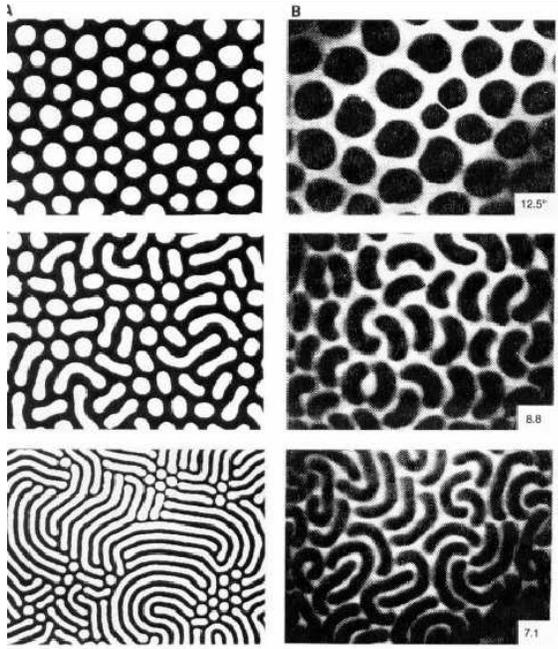}
\vspace*{-0.0cm}
\label{andelman}
\caption{Reproduced with permission from Science, Ref.\cite{Seul}.
Reversible ``strip out'' instability in magnetic
and organic thin films. Period ($L_{D}$) reduction
under the constraint of fixed overall composition
and fixed number of domains leads to elongation
of bubbles. Left panel (A) in magnetic garnet films, this is 
achieved by raising the temperature [labeled in 
(B) in degrees Celsius] along the symmetry axis, $H=0$ (period
in bottom panel, $\sim 10 \mu m$) (see Fig. 5). 
Right panel 
(B) In Langmuir films composed of phospholipid dimyristolphosphatidic
acid (DMPA) and cholesterol (98:2 molar ratio, pH 11), 
this is achieved by lowering the temperature at constant
average molecular density [period in bottom panel,
$\sim 20 \mu m$].}
\end{figure}

In the current article, we focus on translationally invariant 
systems harboring
several interactions of different ranges. To avoid
many of the subtleties related to pure 
long range interactions, we will (unless stated 
otherwise) examine situations wherein screening
is present. With these ingredients in place, 
we will find that the correlation functions
of systems with screened long range interactions
exhibit, in the exactly solvable
large n limit, a peculiar- rather universal- 
divergence of correlation lengths at high temperatures
in many such systems. Notwithstanding this effect,
none of the standard correlation functions
exhibit any pathology- all correlators decay monotonically 
with increasing temperatures. In many such systems,
there are new emergent modulation
lengths governing the size
of various domains. We find that these 
modulation lengths often also
adhere to various scaling laws,
sharp crossovers and divergences
at various temperatures (with no 
associated thermodynamic transition).
We also find that in such systems, correlation
lengths generically evolve into modulation
lengths (and vice versa) at various temperatures.
The behavior of correlation and modulation lengths
as a function of temperature will afford us with certain
selection rules on the possible underlying microscopic
interactions. In their simplest incarnation, for 
systems hosting two competing interactions, our central results
are two fold:  
(i) In canonical systems harboring competing short and 
long range interactions, modulated patterns appear
whose characteristic modulation length is minimal within 
the ground state and slowly slowly increases as
as temperature is raised. The modulation length $L_{D}$
associated with these patterns diverges at a crossover 
temperature $T^{*}$
above which a uniform phase with multiple
correlation lengths appears. The largest correlation
length monotonically increases even as $T \to \infty$
although the prefactor associated with these
correlation rapidly diminishes. 
(ii) By sharp contrast, in systems with only finite range interactions,
the system exhibits a constant number of modulation and 
correlation lengths at all temperatures. Furthermore, in these
systems, canonically, modulations (if they transpire) span
a maximal length scale within the ground state and in the case
of two competing interactions, the modulation $L_{D}$
length decreases as the temperature is raised.
Armed with these general characteristics, we may easily
discern the viable microscopic interactions 
(exact or effective) which underlie temperature
dependent patterns such as those displayed in 
Fig.(\ref{andelman}). Taken at face value,  
our results on the modulation lengths would suggest 
that any two component interaction
theories underlying panel A of Fig.(1) may involve a confluence of 
long and short range interactions
whereas those underlying panel B might involve
only effective short range interactions. 
This may be said without knowing,
a priori, the detailed microscopic
interactions driving these non-uniform
patterns. The simple treatment 
presented below does not account for
the curvature of bubbles and the like.
These may be easily augmented by
inspecting energy functionals
(and their associated free energy extrema) of
various continuum field morphologies under the 
the addition of detailed domain wall tension forms-
e.g. explicit line integrals along the perimeter
where surface tension exists-
and the imposition of additional constraints
via Lagrange multipliers.

\section{Correlation Functions in the large N limit-
general considerations}

All of the results reported in this article were computed
within the spherical or large $n$ limit \cite{kac}. We
focus on translationally invariant systems
whose Hamiltonian is given by 
\begin{eqnarray}
H = \frac{1}{2} \sum_{\vec{x},\vec{y}}
V(\vec{x},\vec{y})S(\vec{x}) 
 S(\vec{y}). 
\label{Ham}
\end{eqnarray} 
Here, the fields $S(\vec{x})$ may portray classical spins
or bosonic fields. The sites $\vec{x}$ and $\vec{y}$
lie on a hypercubic lattice of size $N$ of unit lattice constant. 
In what follows,
$v(\vec{k})$ and $S(\vec{k})$ will denote the Fourier 
transforms of $V(\vec{x}- \vec{y})$ and $S(\vec{x})$.
For analytic interactions, $v(|\vec{k}|)$ is a function of
$k^{2}$ (to avoid branch cuts). 
The spins satisfy a single global spherical constraint,
\begin{eqnarray}
\sum_{\vec{x}} \langle S^{2}(\vec{x}) \rangle = N 
\label{con}
\end{eqnarray}
enforced by a Lagrange multiplier $\mu$
which renders the model quadratic
(as both Eqs.(\ref{Ham}, \ref{con}) are) 
and thus solvable, see e.g. \cite{us}.
In the below we report the results
for classical fields; the results for
bosonic systems are qualitatively the 
same with additional Matsubara frequency
summations in tow.

For our purposes, it suffices to 
note that the two spin correlator,
\begin{eqnarray}
G(\vec{x}) \equiv \langle S(0) S( \vec{x}) \rangle = k_{B} T
\int \frac{d^{d}k}{(2 \pi)^{d}} \frac{e^{i\vec{k} \cdot \vec{x}}}
{v(\vec{k}) + \mu},
\label{corr}
\end{eqnarray}
with $d$ the spatial dimension. To complete the 
characterization of the correlation functions at different 
temperatures, we note that the Lagrange multiplier $\mu(T)$  
is given by the implicit equation $1= G(\vec{x} =0)$ (Eq.(\ref{con})
fused with translational invariance). 
We now investigate the general character of the correlation
functions given by Eq.(\ref{corr}) for rotationally invariant
systems. If the minimum (minima) of
$v(|\vec{k}|)$ occur(s) at momenta $q$ far from
the Brillouin zone boundaries of the cubic lattice then we may set the range
of integration in Eq.(\ref{corr}) to be unrestricted. 
The correlation function is then dominated by the location
of the poles (and/or branch cuts) of $[v(k) + \mu]$. 
Specifically, if $k^{s}[v(\vec{k}) + \mu]$, 
with $s$ an integer, is a polynomial 
\begin{eqnarray}
P(z) = \sum_{m=0}^{M} a_{m} z^{m}
\end{eqnarray}
in $z=k^{2}$ then, upon insertion into Eq.(\ref{corr})
we will find that the correlators generally display a net of 
$M$ correlation and modulation lengths. At very special temperatures,
the Lagrange multiplier $\mu(T)$ may be such that several poles
degenerate into one- thus lowering the number
of correlation/modulation lengths at those special
temperatures. It is important to emphasize that
this multiplicity of roots and thus of correlation/modulation
lengths occurs generally for any $v(k)$ for which 
$P(z)$ is a polynomial of degree $M  \ge 2$- 
multiple length scales appear irrespective
of any competing interactions (alternating or uniform signs 
in $k^{s} v(k)$). What underlies
multiple length scales is the existence of terms
of different ranges (different powers of $z$ in the illustration
above)- not frustration.

\section{General preliminaries: Short and long range interactions}

A screened Coulomb interactions of screening 
length $\lambda$ (i.e. a two spin potential 
$V = \frac{1}{8 \pi} \frac{1}{|\vec{x} - \vec{y}|}
e^{-\lambda |\vec{x} - \vec{y}|}$
in $d=3$, and $V = \frac{1}{4 \pi} 
e^{-\lambda |\vec{x} - \vec{y}|}\ln |\vec{x}- \vec{y}|$
in $d=2$) has the continuum Fourier transformed interaction
kernel $v(k) = [k^{2} + \lambda^{-2}]^{-1}$. 
On a hypercubic lattice, the nearest neighbor interactions in real space
have the lattice lattice Laplacian
\begin{eqnarray} 
\Delta(\vec{k}) = 2 \sum_{l=1}^{d} (1-\cos k_{l}) 
\label{Laplace0}
\end{eqnarray}
as their Fourier transform.
The real lattice Laplacian
\begin{equation}
  \langle \vec{x}| \Delta |\vec{y} \rangle  = \left\{ \begin{array}{ll}
      2d & \mbox{ for $\vec{x}=\vec{y}$} \\
      -1 & \mbox{ for $|\vec{x}-\vec{y}| = 1$}
\end{array}
\right.
\label{Laplace1}
\end{equation}
Notice that  $\langle \vec{x}| \Delta^{R} |\vec{y} \rangle = 0 
\mbox{ for $ |\vec{x}-\vec{y} |> R$}$,
where $R$ is the spatial range over which the interaction kernel
is non-vanishing. Eq.(\ref{Laplace1}) corresponds to a system is of Range=2,
\begin{eqnarray}
  \langle \vec{x}| \Delta^{2}| \vec{y} \rangle \mbox{ } 
= \mbox{ }&& 2d(2d+2) \mbox{
    for } \vec{x} = \vec{y} \nonumber
  \\
  \ \ && -4d \mbox{ for  } |\vec{x}-\vec{y}| = 1 \nonumber \\
  \ \ && 2 \mbox{ for } (\vec{x} -\vec{y}) = 
(\pm \hat{e}_{\ell} \pm \hat{e}_{\ell^{\prime}}) \mbox{ where  }
 \ell \neq \ell^{\prime}      \nonumber \\
  \ \ && 1 \mbox{ for a $ \pm 2 \hat{e}_{\ell} $ separation}.
\label{Laplace2}
\end{eqnarray}
In the continuum (small $k$) limit, 
$\Delta \to z = k^{2}$. 

For simplicity, many of
the examples which we will employ
to illustrate the general premise
of the behavior of correlations
in systems hosting long and short range
interactions, the kernel $v(\vec{k})$
(the Fourier transform of $V(\vec{x},\vec{y})$
of Eq.(\ref{Ham})) will, in the continuum
limit, be a simple function of $k^{2}$
or of $\Delta$ whenever the finite lattice
constant is kept. 

Simple effects of tension may be emulated via
a $g(\nabla \phi)^{2}$ term in the Hamiltonian
where $\phi$ is a constant
in a uniform domain. Upon Fourier transforming,
such squared gradient terms lead to an effective $k^{2}$ in the 
$\phi$ space kernel. Similarly,
the effects of curvature notable in many mixtures and 
membrane systems are
often captured by terms involving
$(\nabla^{2} h)$ with $h$ 
a variable parameterizing the profile;
at times the interplay of such curvature terms
with others leads, in the aftermath, to 
a simple short range $k^{4}$ term in the interaction
kernel. An excellent review of 
these issues is addressed in 
\cite{Seul}. Screened dipolar interactions
and others may be easily emulated
by terms such as $(k^{2} + \lambda^{2})^{-p}$ 
with $p>0$.

\section{multiple range interactions}

We now summarize the situation wherein 
two dominant interactions compete whenever one of
the interactions is of infinite range (albeit
being screened) while the other is of 
finite range (i.e. $V$ strictly vanishes
for separations $|\vec{x} - \vec{y}| > R$). 
Such situations arise in many
systems \cite{Seul}.

{\bf 1})  The high temperature correlation functions 
are, in many instances, sum of several exponential 
pieces; e.g. the correlation function
\begin{eqnarray}
\langle S(\vec{x}) S(\vec{y}) \rangle = \frac{1}{|\vec{x} - \vec{y}|^{d-2}}
 \Big( A_{1}e^{- |\vec{x} - \vec{y}|/\xi_{1}} \nonumber
\\ + A_{2} 
e^{- |\vec{x} - \vec{y}|/\xi_{2}} + ... \Big).
\label{A12}
\end{eqnarray} 
[At least one of the correlation lengths ($\xi_{i}$)
diverges.]

{\bf 2}) At low temperatures, translationally 
frustrated systems with competing interactions 
on different length scales
display modulations (i.e. an oscillatory
spatial dependence of the 
correlation functions),
e.g. 
\begin{eqnarray}
\langle S(\vec{x}) S(\vec{y}) \rangle 
\sim \frac{\cos(p|\vec{x} - \vec{y}|)}{|\vec{x} - \vec{y}|^{d-2}} 
e^{- |\vec{x} - \vec{y}|/\xi} + ...
\end{eqnarray}

{\bf 3}) When present in systems with frustrating long-range interactions
(e.g. the Coulomb Frustrated Ferromagnet originally introduced
to portray frustrated charge separation in the cuprates, 
\cite{us}, \cite{zohar}, \cite{Low}) 
for which, in Eq.(\ref{Ham}), $v(k) = k^{2} + Q k^{-2}$, 
the modulation lengths monotonically increase
with increasing temperatures. In the specific case of
the Coulomb frustrated ferromagnet, the modulation length
$(2 \pi/ p)$ monotonically increases with temperatures
until it diverges at a disorder line temperature $T^{*}$\cite{zohar}. 
We find that the near this temperature (i.e. for $T = T^{*~-}$), 
the modulation length scales as $(T^{*}-T)^{-1/2}$. 
It should be emphasized that this divergence notwithstanding,
the system does not exhibit a phase transition at $T^{*}$.
The free energy is analytic at this temperature. Nevertheless, 
the free energy can be made to have a singularity at 
$T = T^{*}$ if the long range interaction is turned 
off ($Q=0$)- this allows us to view the divergence of the modulation 
length as sparked by a critical point which is 
``avoided''. \cite{us},\cite{zohar}

{\bf 4}) In most systems, the 
sum of the number of correlation and the number of modulation lengths
is conserved as a function of temperature 
apart from special crossover temperatures. 
In the example of the Coulomb Frustrated 
Ferromagnet, at temperatures higher than this crossover temperatures
$T> T^{*}$ two correlation lengths appear as in (1). After
diverging at $T=T^{*}$ the modulation length turns into a correlation
length at higher temperatures. 
All crossovers may be traced by examining the
dynamics of the poles and branch cuts of $1/[v(k) + \mu]$ 
as the temperature (implicit in $\mu$) is varied. The merger of
two or more poles at special temperatures 
leads to a temporary annihilation of one (or more length)
which is generically restored as the temperature is 
continuously varied. 

{\bf 5}) In systems with only two finite range interactions
(of different scales), the modulation length monotonically
decreases with increasing temperatures. This, combined with
(3), affords a stringent selection rule on the viable interactions
underlying various experimentally observed modulation lengths. 
In canonical systems harboring only 
finite range interactions, the number of
correlation lengths and the number of modulation lengths
are both independently conserved at all temperatures
(much unlike the case for long range interactions
where only the sum of the two is conserved).

Although these characteristics are general, it is useful to
provide expressions for specific cases. In what follows, 
we briefly sketch the behavior when two interactions
of two different ranges, appear in unison. 
In section(\ref{Coul+}), we discuss the case
of an infinite range interaction (wherein $\langle \vec{x}| V
| \vec{y} \rangle \neq 0$ for all $\vec{x} \neq \vec{y}$)
screened Coulomb interaction existing side
by side with a short range nearest neighbor 
exchange interaction. We follow, in section(\ref{multiple})
by an investigation of a system hosting several 
finite range interactions. We the consider, in section(\ref{longrangeint}), 
with brief remarks concerning the behavior in
the case of a system in which long range interactions 
compete. General remarks concerning
the possibility of first order transitions (section(\ref{1storder}))
in the modulation length and a rather 
universal domain length exponent (section(\ref{exponent}))
conclude the article.  

\section{Coexisting short and long range interactions}
\label{Coul+}

We begin with the ``Screened Coulomb Ferromagnet'',
capturing the competition between a 
repulsive screened Coulomb interactions
and short range attractions.
Here, the Fourier transform of the interaction kernel of Eq.(\ref{Ham}) 
is
\begin{eqnarray}
v(k) = k^{2} + \frac{Q}{k^{2} + \lambda^{2}}
\label{Yukawa+}
\end{eqnarray}
wherein a screened Coulomb interaction competes
with a short range ferromagnetic interaction.
Not too surprisingly, such an kernel
naturally appears in many systems
e.g. screened models of frustrated phase
separation in the cuprates \cite{steve}.
Here, at high temperatures ($T> T^{*}$ wherein a $T^{*}$
is given by $\mu(T^{*}) = \lambda^{2}+ 2 \sqrt{Q}$),
the pair correlator in dimension $d=3$
\begin{eqnarray}
G(\vec{x}) = \frac{k_{B}T}{4 \pi |\vec{x}|} \frac{1}{\beta^{2} - \alpha^{2}} 
\nonumber
\\ \times
[e^{- \alpha |\vec{x}|} (\lambda^{2} - \alpha^{2}) - e^{-\beta |\vec{x}|}
(\lambda^{2} - \beta^{2})].
\label{G1}
\end{eqnarray}
Here, 
\begin{eqnarray}
\alpha^{2}, \beta^{2} = 
\frac{\lambda^{2}+ \mu \mp \sqrt{(\lambda^{2} - \mu)^{2} - 4 Q}}{2}.
\label{3DHighT}
\end{eqnarray}
For $T<T^{*}$, 
\begin{eqnarray}
G(\vec{x}) = \frac{k_{B}T}{8 \alpha_{1} \alpha_{2} \pi |\vec{x}|} 
e^{-\alpha_{1} |\vec{x}|} \nonumber
\\ \times
[(\lambda^{2} - \alpha_{1}^{2} + \alpha_{2}^{2}) \sin \alpha_{2}|\vec{x}|
+ 2 \alpha_{1} \alpha_{2} \cos \alpha_{2} |\vec{x}|]
\label{G2}
\end{eqnarray}
where $\alpha = \alpha_{1} + i \alpha_{2} = \beta^{*}$.

Similarly, in $d=2$, for $T>T^{*}$,
\begin{eqnarray}
G(\vec{x}) = \frac{k_{B}T}{2 \pi} \frac{1}{\beta^{2} - \alpha^{2}}
[(\lambda^{2}- \alpha^{2}) K_{0}(\alpha |\vec{x}|) \nonumber
\\ + (\beta^{2} - \lambda^{2})
K_{0}(\beta |\vec{x}|)],
\label{2DHighT}
\end{eqnarray}
with the Bessel function
$K_{0}(x) = \int_{0}^{\infty} dt \frac{\cos xt}{\sqrt{1+ t^{2}}}$.
Much as in the three dimensional case, the high temperature
correlator may be analytically continued to temperatures $T<T^{*}$.

\begin{figure}
\centering
\hspace{0.0in}{\psfig{figure=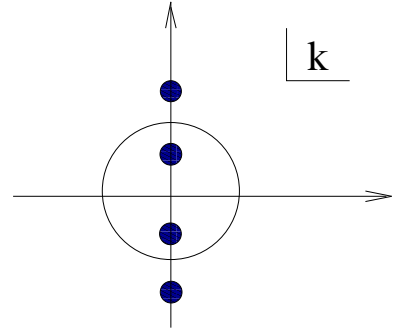,width=2.0in,clip=}}
\caption{The location of the poles in the complex k plane 
at high temperatures for competing long range-short range interactions
(e.g. for temperatures $T>T^{*}$ the model of Eq.(\ref{Yukawa+}) 
describing screened Coulomb repulsions competing
with short range attractions).}
\label{fig:poles1}
\end{figure}

\begin{figure}
\centering
\hspace{0.0in}{\psfig{figure=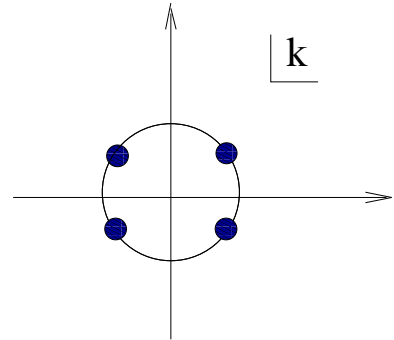,width=2.0in,clip=}}
\caption{The location of the poles at low temperatures
for competing long range-short range interactions (e.g. 
the system of Eq.(\ref{Yukawa+}) at temperatures
$T_{c}< T <T^{*}$). At $T$ is lowered the poles approach from 
both sides the real axis (the inverse correlation length-
the height along the imaginary k axis- decreases. 
At $T_{c}$, the poles unite in pairs
on the positive and negative portions of the real axis.}
\label{fig:poles2}
\end{figure}

We alert the reader that two correlation lengths appear for $\mu^{2}>4Q$ 
({\em including all unfrustrated screened attractive 
Coulomb ferromagnets} (those with $Q<0$)).
The evolution of the correlation functions may be traced by 
examining the dynamics of the poles in the complex $k$ plane
as a function of temperature. At high temperatures, 
the correlation functions are borne
by poles lying on the imaginary axis. In the high temperature
limit ($T \to \infty$), one of the poles tends to $k=0^{+}$
leading to a {\em divergent correlation length}! No paradoxes
are, however, encountered in this limit
as the prefactor associated with this 
correlation length (as seen in Eqs.(\ref{3DHighT}, \ref{2DHighT}) 
tends to zero as $T \to \infty$ and all correlations
decay monotonically with increasing temperature. At $T=T^{*}$ the poles
merge in pairs at $k= \pm i \sqrt{\lambda^{2}+ \sqrt{Q}}$.
Henceforth, at lower temperatures, the poles move off the imaginary axis
(leading in turn to oscillations in the correlation
functions). The norm of the poles, $|\alpha| = 
(Q+ \lambda^{2} \mu(T))^{1/4}$, becomes constant
in the limit of vanishing screening ($\lambda =0$)
wherein the after merging at $T=T^{*}$, the
poles slide along a circle. In the low temperature
limit of the unscreened Coulomb ferromagnet,
the poles hit the real axis at finite $k$,
reflecting oscillatory modulations in the 
ground state. In the presence of 
screening, the pole trajectories are slightly
skewed yet for $Q> \lambda^{4}$,
$\alpha$ tends to the ground state
modulation wavenumber $\sqrt{\sqrt{Q} - \lambda^{2}}$. 
If the screening is sufficiently large,
i.e. if the screening length is shorter than 
the natural period favored by a balance between
the unscreened Coulomb interaction and the
nearest neighbor attraction ($\lambda > Q^{1/4}$),
then the correlation functions never exhibit 
oscillations. In such instances, the poles continuously
stay on the imaginary axis and, at low temperatures,
one pair of poles veers towards $k=0$ reflecting
the uniform ground state of the heavily screened
system. In Figs.(\ref{fig:poles1},\ref{fig:poles2}),
the evolution of the pole locations in traced at different
temperatures $(T>T^{*}, T<T^{*})$ in the limit of weak screening.
To summarize the observed physics at hand, at high temperatures
$G(x)$ is a sum of two decaying exponentials (one of 
which has a correlation length which diverges in 
the high temperature limit). For $T<T^{*}$ in under-screened
systems, one of the correlation lengths transforms into
a modulation length characterizing low temperature oscillations.
At the cross-over temperature $T^{*}$, the modulation length
is infinite. As the temperature is progressively lowered,
the modulation length decreases in size- until it
reaches its ground state value. Connoisseurs
may recognize $T^{*}(Q,\lambda)$ as a 
``disorder line'' like temperature.

An analytical thermodynamic crossover does occur
at $T=T^{*}$. A large $n$ calculation of the 
free energy via equipartition reveals that 
the internal energy per partice
\begin{eqnarray}
\frac{U}{N} = \frac{1}{2}(k_{B}T- \mu),
\label{Umu}
\end{eqnarray} 
To ascertain a crossover in $U$ and that in
other thermodynamic functions, the 
forms of $\mu$ both above and below $T^{*}$
may be easily gleaned from the spherical normalization
condition to find that the real valued functional 
form of $\mu(T)$ changes \cite{explainlong}.

We note, in passing, that the system orders
at $T=T_{c}$ given
by 
\begin{eqnarray}
\frac{1}{k_{B}T_{c}} = \int \frac{d^{d}k}{(2 \pi)^{d}} 
\frac{1}{v(\vec{k}) - v(\vec{q})}.
\label{tg}
\end{eqnarray}
Here, for $Q > \lambda^{4}$, the modulus of the 
minimizing (ground state) wavenumber ($|\vec{q}|$) is
given by 
\begin{eqnarray}
q = \frac{2 \pi}{L_{D}^{g}} = \sqrt{\sqrt{Q} - \lambda^{2}},
\end{eqnarray}
with $L_{D}^{g}$ the ground state modulation length.
Associated with this wavenumber is the kernel
$v(\vec{q}) = 2 \sqrt{Q} - \lambda^{2}$
to be inserted in Eq.(\ref{tg}) for an evaluation of the
critical temperature $T_{c}$.
Similarly, the ground state wavenumber 
$\vec{q} =0$ whenever
$Q < \lambda^{4}$.
Needless to say, whenever $Q > \lambda^{4}$
and modulations transpire for temperatures $T<T^{*}$, the
critical temperature at which 
the chemical potential of Eq.(\ref{corr}),
$\mu(T_{c}) = \lambda^{2} - 2 \sqrt{Q}$,
is lower than the crossover temperature
$T^{*}$ (given by $\mu(T^{*}) = \lambda^{2} + 2 \sqrt{Q}$)
at which modulations first start to appear.
The Yukawa Ferromagnet is found to have
$T_{c}(Q=\lambda^{4})>0\mbox{ in $ d>4$ }$ and in any dimension
$T_{c}(Q>\lambda^{4})=0$. 
For small finite $n$ a first order Brazovskii transition
may replace the continuous transition occurring at $T_{c}$
within the large $n$ limit \cite{Braz}. Depending
on parameter values such an equilibrium transition may or may not
transpire before a possible glass transition may occur \cite{peter}. 
 
\section{Multiple short range interactions}
\label{multiple}

Our central thesis is that in many 
models in which 
short range 
interactions compete with one another,
the modulation length always 
varies with temperature. This is
in contrast to 
in systems with infinite range
interactions wherein the modulation 
length increases as temperature
is raised.

We now illustrate this general premise 
by a few examples: (i) With the conventions
of the Fourier transformed kernel of 
Eq.(\ref{Ham}) and Eqs.( \ref{Laplace0}, \ref{Laplace1}),
the Fourier transformed interaction kernel
\begin{equation}
  v(\vec{k}) = [\Delta^{3}-\Delta_{0}^{3}]^{2}
\end{equation}
corresponds to a finite range interaction
linking sites which are, at most,
six lattice units apart.
In what follows we employ the shorthand
$ c \equiv (\mu -\mu_{min})$ with $\mu_{\min} = \Delta_{0}^{6}$. 
The poles, $\{\Delta_{\alpha}^{\pm} \}_{\alpha=1}^{3}$, of the correlator
$G(\vec{k}) = k_{B}T [c+v(\vec{k})]^{-1}$ are given by
\begin{eqnarray} 
[\Delta_{0}^{3} \pm \imath
  \sqrt{c}]^{1/3} \exp{(2 \pi \alpha \imath /3)}.
\end{eqnarray}

In the complex plane, the poles
$\{\Delta_{\alpha}^{\pm}\}$ lie on the vertices of a hexagon. 
There exist, at least, two poles with different
values of $|Im\{k_{i}\}|$ leading to, at least, 
two different correlation lengths, 
$\xi_{i} = |Im\{k_{i}\}|^{-1}$ at all temperatures 
(and apart from special degenerate usually to three
correlation lengths). The existence 
of multiple correlation lengths $\{\xi_{i}\}$ 
are the rule in systems of multiple range interactions- 
including unfrustrated systems with no competing interactions.
Their presence (as well as the existence of
multiple modulation lengths) enriches the
scope of possible scaling functions,
allowing us to construct a greater multitude
of scaling functions $F(\Gamma; \{\xi_{i}\}, \{L_{D;i}\})$, 
with $\Gamma$ a set of external parameters, than those 
present in standard critical systems where only
one length (a single correlation length)
sets the scale. For a finite ranged interaction kernel 
$V$ which is a general polynomial of the lattice Laplacian
$\Delta$, the system possesses several
modulation lengths and several correlation lengths at all temperatures;
barring few exceptions - their net number is conserved as temperature is
varied.  Generically, for a finite ranged interaction which
is a polynomial in the Lattice Laplacian $\Delta$,
the system possesses a fixed number of correlation lengths
and a fixed number of modulation lengths; there is no
sharp analogue of $T_{i}$ wherein modulation lengths turn into
correlation lengths. Such multiple correlation lengths often present 
for such kernels $v(\vec{k})$ which are functions of the lattice Laplacian
$\Delta$ are accompanied by a $T_c$ discontinuity (an ``avoided
critical point'' \cite{us}, \cite{zohar}) 
in sufficiently high dimension when appropriate competition is present to
allow real roots $0 < \{\Delta_{i}\} < 4d $ when $\mu =\mu_{min}$.
All cross-over temperatures [$T_{i=1,2,...,p}$]
(including low dimensional systems which possess no critical behavior
for zero frustration), at which correlation lengths disappear and turn
into modulation lengths, tend continuously to the avoided critical
temperature (or its analytic continuation for low dimensions - in
high dimensions such an ``avoided critical temperature'' \cite{us}, 
\cite{zohar} becomes critical for zero frustration).
These crossover temperatures [$T_{i=1,2,...,p}$] are more dramatic 
for non-analytic functions of $\Delta$ such as the
frustrated screened Coulomb interaction investigated in
Section(\ref{Coul+}), wherein 
the domain length diverges.

(ii) We next consider a very prototypical
interaction kernel appearing, amongst others, 
in amphiphile problems and crumpled
elastic membrane systems, see e.g. \cite{Seul}. 
For the short-range (Teubner-Strey) correlator
\begin{eqnarray}
G^{-1}(\vec{k}) = a_{2}  +c_{1}k^{2}+c_{2}k^{4},
\end{eqnarray}
it is a simple matter to show that
\begin{eqnarray}
G(\vec{x}) \sim \frac{\sin \kappa x}{\kappa x} \exp[-x/\xi],
\end{eqnarray}
where
\begin{eqnarray}
\kappa = \sqrt{\frac{1}{2}
\sqrt{\frac{a_{2}}{c_{2}}}- \frac{c_{1}}{4c_{2}}} \nonumber
\\ \xi^{-1} = \sqrt{\frac{1}{2}\sqrt{\frac{a_{2}}{c_{2}}}
+ \frac{c_{1}}{4c_{2}}}.
\end{eqnarray}

In amphiphilic systems $a_{2}$ and $c_{1,2}$ are functions
of amphiphile concentration (as well as temperature).

The above two examples reaffirm out claim
that in many simple
thermodynamical models,
short range interactions,
the modulation length (if it exists)
typically increases as 
the temperature is lowered.
The converse typically occurs
in systems hosting competing
infinite range and finite 
range interactions. 

When, in the notation of the Fourier transformed 
kernel of Eq.(\ref{Ham}), $v(\vec{k}) = k^{4}$,
\begin{eqnarray}
G(\vec{x}) = -\frac{1}{4 \pi x \sqrt{\mu}} \exp{[-x/\xi]} \sin \kappa x.
\end{eqnarray}
Thus within the spherical model of this 
range two system with a sole overall
ferromagnetic interaction
($v(\vec{k}) = k^{4}$) will
display thermally induced oscillations.
At $T=0$ the (ferromagnetic)
ground state is unmodulated.

\section{Multiple Long Range Interactions}
\label{longrangeint}

A different behavior is seen when two long range interactions
compete (e.g. $v(k) = Ak^{-2} + Bk^{-4}$ with $A<0$ and $B>0$).
In such instances, modulations are present at all temperatures.
Moreover, by sharp contrast to the competing 
long-range and short range interactions investigated
earlier, within the large $n$ limit, the modulations 
become more and more acute
with a length which tend to zero in the 
high temperature limit. 

\section{First order transitions in the modulation 
length}
\label{1storder}

In the examples furnished above and several 
of our general maxims in the introduction,
we focused on systems in which only two interactions
of different ranges exist in a single system. 
In these systems, we found within the large $N$ limit
that the modulation lengths were always monotonic
in temperature. Needless to say, this need not be 
the case yet within the large $n$ limit this generally 
requires the existence of interactions spanning more ranges.
The ground state modulation lengths
(the reciprocals of Fourier modes $\{\vec{q}_{i}\}$
minimizing the interaction
kernel) need not be 
continuous as a function
of the various parameters: a ``first order transition''
in the value of the ground
state modulation lengths
can occur. Such a possibility is quite obvious and 
need not be expanded upon in depth.
Consider, for instance, the Range=3 interaction kernel
\begin{eqnarray}
v(\vec{k}) =  a [\Delta+ \epsilon] + \frac{1}{2} b [\Delta+\epsilon]^{2} 
+ \frac{1}{3} c [\Delta + \epsilon]^{3}, 
\end{eqnarray}
\newline
with [$  0< \epsilon \ll 1$] and $c>0$.  
If $a >0$ and $b<0$, then there are three
minima, i.e.  $[\Delta+\epsilon] =0$ and $[\Delta+\epsilon] = 
\pm m_{+}^{2}$ 
where $m_{+}^{2} = \frac{1}{2c} [ -b+ \sqrt{b^{2}-4ac}]$.
the locus of points in the $ab$ plane where the three minima are equal
is determined by
$v(\vec{k}) =0$, which leads to 
$m_{+}^{2} = -\frac{4a}{b}$.
Thus, $b= -4 \sqrt{ca/3}$
is a line of ``first order transitions'',in which the minimizing 
$[\Delta+ \epsilon]$ (and thus the minimizing wavenumbers)
changes discontinuously by an amount
$\Delta m = (- \frac{4a}{b})^{1/2} = (\frac{3a}{c})^{1/4}$.

\bigskip

\section{A Universal Domain length exponent}
\label{exponent}

Invoking the normalization condition $G(\vec{x}=0) =1$
in Eq.(\ref{corr}), 
we find that given the competing screened 
Coulomb and short range attraction of 
Eq.(\ref{Yukawa+}) and more general
kernels $v(\vec{k})$ in which infinite-range interactions 
augment finite ones, in any problem in
which the crossover temperature 
$T^{*}$ is finite, the modulation length 
\begin{eqnarray}
L_{D} \sim (T^{*}-T)^{-\nu_{L}},
\end{eqnarray} 
with the (large $n$) domain length exponent $\nu_{L} = 1/2$
in any dimension $d$.

\section{Conclusions}

In conclusion, our major finding is 
a general evolution of 
modulation and correlation 
lengths as a function of temperature
in different classes of systems, those
harboring infinite range (including screened)
interactions and those having only 
short range interactions. These ``selection rules''
impose constraints on candidate theories 
describing a system in which the 
empirical behavior of modulation 
lengths is known. We further 
elucidated on the peculiar thermal evolution of the multiple correlation 
lengths in many systems having long range interactions,
the largest of which may increase monotonically
in temperature (even as $T \to \infty$).

It is a pleasure to acknowledge many discussions 
with David Andelman, Lincoln Chayes, 
Daniel Kivelson, Steven Kivelson,
Joseph Rudnick, Gilles Tarjus, and Peter
Wolynes.

* Current address: Theoretical Division,
Los Alamos National Laboratory, Los Alamos, NM 87545, USA


\begin{references}



\bibitem{long} T. Dauxios, S. Ruffo, E. Arimondo, M. Wilkens (Eds.),
``Dynamics and Thermodynamics of Systems with Long Range Interactions'',
Lecture Notes in Physics, {\bf 602}, Springer (2002)

\bibitem{lilly}
M. P. Lilly, K. B. Cooper, J. P. Eisenstein, L. N. Pfeiffer, and K. W. West
Phys. Rev. Lett. {\bf 82}, 394 (1999) 
(cond-mat/9808227)

\bibitem{QHE}
E. Fradkin, S. A. Kivelson
Phys. Rev. B {\bf 59}, 8065 (1999)
(cond-mat/9810151)

\bibitem{Fogler}
M. M. Fogler, cond-mat/0111001, 
p. 98-138, in High Magnetic Fields: Applications in Condensed Matter Physics 
and Spectroscopy, ed. by C. Berthier, L.-P. Levy, G. Martinez 
(Springer-Verlag, Berlin, 2002)

\bibitem{amp} Hyung-June Woo, C. Carraro, D. Chandler, Phys. Rev. E, 
{\bf 52}, 6497 (1995); F. Stilinger, J. Chem. Phys. {\bf 78}, 4655 
(1983); L. Leibler, Macromolecules, {\bf 13}, 1602 (1980);
T. Ohta and K. Kawasaki, Macromolecules, {\bf 19}, 2621 (1986)  

\bibitem{vortex3} H. Kleinert, ``Gauge Fields in Condensed Matter'',
World Scientific (1989), volume II 

\bibitem{vortex1} P. H. Chavanis, ``Statistical Mechanics of Two Dimensional
Vortices and Stellar systems'' in Ref.[\cite{long}]. 

\bibitem{vortex2} H. Kleinert, ``Gauge Fields in Condensed Matter'',
World Scientific (1989), volume I 

\bibitem{Seul}
M. Seul and D. Andelman, Science {\bf 267}, 476 (1995)

\bibitem{wavpart} Y. Elskens, ``Kinetic Theory for Plasmas and Wave-particle
Hamiltonian Dynamics'', in \cite{long}; Y. Elskens and D. Escande, 
``Microscopic Dynamics of Plasmas and Chaos'', IOP publishing, Bristol (2002)





\bibitem{steve}
V. J. Emery and S. A. Kivelson, Physica C {\bf 209}, 597 (1993)

\bibitem{us}
L. Chayes et al. 
Physica A {\bf 225}, 129 (1996) 

\bibitem{zohar}
Z.~Nussinov et al., Phys. Rev. Letters 83, 
472 (1999)

\bibitem{Low}
U. L\"ow et al.,
Phys. Rev. Lett. {\bf 72}, 1918 (1994) 

\bibitem{carlson}
E. W. Carlson, V. J. Emery, S. A. Kivelson, D. Orgad
cond-mat/0206217, Review chapter to appear 
in `The Physics of Conventional and Unconventional 
Superconductors' ed. by K. H. Bennemann and J. B.
       Ketterson (Springer-Verlag)

\bibitem{cheong}
S- W. Cheong et al.,
Phys. Rev. Lett. {\bf 67}, 
1791 (1991)

\bibitem{golosov}
D. I. Golosov, Phys. Rev. B, vol. 67, 064404 (2003)
(cond-mat/0206257)



\bibitem{dk}
D. Kivelson et al.
(Physica A {\bf 219}, 27  (1995))


\bibitem{peter} J. Schmalian and P. G. Wolynes, Phys. Rev. Lett. {\bf
85}, 836 (2000); H. Westfahl, Jr., J. Schmalian, and P. G.
Wolynes, Phys. Rev. B {\bf 64}, 174203 (2001)

\bibitem{Gilles}
M. Grousson et al., Phys. Rev. E {\bf 62}, 7781 (2000)


\bibitem{new} Z. Nussinov, Phys. Rev. B,  {\bf 69}, 014208 (2004)

\bibitem{der}B. V. Derjaguin and L. Landau, Acta Physiochim, URSS 
{\bf 14}, 633 (1941); E. J. Verwey and J. T. G. Overbeek 
{\em Theory of Stability of Lyophobic
Colloids} (Elsevier, Amsterdam, 1948)


\bibitem{reich} C. Reichhardt and C. J. Olson, Phys. Rev. Lett., {\bf 88},
248301 (2002)
 


\bibitem{barre} J. Barre, D. Mukamel, S. Rufffo, ``Ensemble inequivalence
in mean field models of magnetism'' in \cite{long}
 
\bibitem{azbel} Mark Ya. Azbel\&rsquo, cond-mat/03020371,
Phys. Rev. E 68, 050901 (2003)

\bibitem{kac}  T. H. Berlin and M. Kac, Phys. Rev. {\bf 86}, 821 (1952);
H. E. Stanley, Phys. Rev. {\bf 176}, no. 2, 718 (1968)
 

\bibitem{explainlong} For instance, 
in three dimensions, with $\Lambda = 2 \pi/a$ 
an ultra-violet cutoff with $a$ the lattice unit length, 
at high temperatures, $T>T^{*}$, this leads to the following
implicit equation for $\mu(T)$ in the case of
the screened Coulomb ferromagnet,
\begin{eqnarray}
\frac{1}{T} = \frac{\Lambda}{2 \pi^{2}} +
\frac{\sqrt{2}}{4 \pi^{2} p}\nonumber
\\ \times \Big( \frac{\lambda^{2} \mu - \mu^{2} 
+ \mu p - 2Q}{\sqrt{\lambda^{2}+ \mu + p}} \tan^{-1} 
(\frac{\Lambda \sqrt{2}}{\sqrt{\lambda^{2} + \mu + p}}) \nonumber
\\ - \frac{\lambda^{2} \mu - \mu^{2} + \mu p + 2Q}
{\sqrt{\lambda^{2} + \mu - p}}
\tan^{-1}(\frac{\Lambda \sqrt{2}}{\sqrt{\lambda^{2}  + \mu - p}}) \Big).
\label{explicitT}
\end{eqnarray}
Here, we employed the shorthand 
$p \equiv \sqrt{(\mu - \lambda^{2})^{2} - 4Q}$.
This parameter $p$ 
vanishes at the crossover temperature $T^{*}$ at which 
a divergent modulation length makes an appearance, $p(T=T^{*})=0$.
At low temperatures, $T<T^{*}$, $p$ becomes imaginary and 
an analytical crossover occurs to another real functional form. 


\bibitem{Braz} S. Brazovskii, Sov. Phys. JETP {\bf 41}, 85 (1975)


\end{references}
\end{document}